\documentclass{ws-ijmpd}

\begin{document}

\markboth{M. Govender}
{Non-adiabatic spherical collapse with a two-fluid atmosphere}

%
\catchline{}{}{}{}{}
%

\title{Non-adiabatic spherical collapse with a two-fluid atmosphere  }

\author{M. Govender\footnote{E-mail: govenderm43@ukzn.ac.za}}

\address{Astrophysics and Cosmology Research Unit, School of Mathematical Sciences, \\
University of KwaZulu--Natal, Private Bag X54001, Durban 4000, South Africa}

\maketitle

\begin{history}
\received{Day Month Year}
\revised{Day Month Year}
\end{history}

\begin{abstract}
In this work we present an exact model of a spherically symmetric star undergoing dissipative collapse
in the form of a radial heat flux. The interior of the star is matched smoothly to the generalised Vaidya line element representing
a two-fluid atmosphere comprising null radiation and a string fluid. The influence of the string density on the thermal behaviour of the model is investigated by employing a causal heat transport equation of Maxwell-Cattaneo form.
\end{abstract}


\section{Introduction}
The first attempt at investigating the end-state of gravitational collapse of a bounded matter distribution within the realm of general relativity was due to the efforts of Oppenheimer and Snyder \cite{snyder} in which they considered the gravitational collapse of a dust sphere.
Although highly idealised, the Oppenheimer-Snyder dust model paved the way for more realistic collapse models. The singularity theorems of Hawking and Penrose assert that given a trapped surface a singularity forms which could be covered or naked \cite{pen,josh}.  The discovery of naked singularities as the possible end-states of gravitational collapse added a new dimension to the study of cosmic censorship. This led to a renewed search of physically reasonable matter distributions avoiding the formation of the horizon during gravitational collapse. The inclusion of pressure, shear, bulk viscous effects and dissipation in the form of heat flow within the stellar fluid became possible to study with the discovery of Vaidya's radiating solution for the exterior of the star \cite{Bonnor,crack,Vaidya}. The junction conditions meshing the interior spacetime to the Vaidya atmosphere were obtained by Santos in 1985 \cite{Santos}. The main feature of the dissipative collapse model is the nonvanishing of the radial pressure at the boundary of the collapsing star. In the presence of a radial heat flux, Santos showed that the pressure is proportional to the magnitude of the heat flux at the surface of the star. This gives rise to a dynamical equation which, in principle, establishes the temporal evolution of the collapsing star. The problem of dissipative collapse was extensively investigated in its generality by Herrera and co-workers who established many general frameworks of the collapse scenario. It was then possible to obtain exact solutions describing radiating, collapsing spheres and these were largely due to Herrera and co-workers and Maharaj and co-workers, amongst others.

The Vaidya solution is the unique spherically symmetric solution of the Einstein field equations representing a radiation atmosphere. This solution has been extensively studied in the literature especially within the realm of cosmic censorship including the
collapse of charged null fluid, mixture of quark and null fluid, higher dimensional radiation fluids and and modified gravity models.  The so-called generalised Vaidya solution in which the mass function depends on both the radial and null coordinate can be thought of as an atmosphere comprising of null fluid and strings. The generalised Vaidya solution has been utilised to study the outcome of gravitational collapse in various scenarios including the formation of naked singularities in classical general relativity, modeling dissipative collapse and studying radiating black hole solutions in higher dimensions within the Einstein-Yang-Mills-Gauss-Bonnet framework\cite{glass1,glass2,Kesh,SG1,SG2}. The junction conditions required for the matching of a spherically symmetric line element to the generalised Vaidya solution was presented by Maharaj {\em et al}\cite{Sunil}. These results generalised the earlier junction conditions obtained by Santos. It was shown that the continuity of the momentum flux across the boundary of the collapsing star required the nonvanishing of the surface pressure. In addition to the pressure being proportional to the magnitude of the heat flux (as in the case of a pure radiation atmosphere), it was shown that for the two-fluid atmosphere the surface pressure depends on the heat flux and the string density.

The aim of this paper is to investigate the influence of the string density on the thermal behaviour of the collapsing star. To this end, we adopt a causal heat transport equation to obtain both the Eckart temperature as well as the causal temperature throughout the stellar interior. This paper is organised as follows: In section 2 we present the interior spacetime and matter content. The generalised Vaidya solution together with the appropriate junction conditions are presented in section 3. The evolution of the temperature profile in both the causal and noncausal theories are obtained in section 4. A discussion of the time of formation of horizon is discussed in section 5. We finally conclude with a discussion of our results and make comparisons with earlier findings in section 6.

\section{Stellar Interior}
The interior spacetime of our radiating star is described by a spherically symmetric, shear-free line element given by\cite{Bonnor}
\begin{equation}
\label{g1} ds^2 =-A^2 dt^2 +B^2[ dr^2 +r^2 (d\theta^2 +\sin^2
\theta d\phi^2)],
\end{equation}
where $A(r, t)$ and $B(r, t)$ are metric functions yet to be determined.  The interior stellar material is represented by a perfect fluid with heat flux:
\begin{equation}\label{g2}
T_{ab} = (\mu + p)u_a u_b + p g_{ab} + q_a u_b
          + q_b u_a
\end{equation}
where $\mu$ is the energy density, $p$ is the pressure and $q =
(q^aq_a)^{\frac{1}{2}}$ is the magnitude of the heat flux. The fluid
four--velocity ${\bf u}$ is comoving and is given by
\begin{equation}
u^a = \displaystyle\frac{1}{A} \delta^{a}_0\label{2'}
\end{equation}
The heat flow vector takes the form
\begin{equation}
q^a = (0, q, 0, 0) \label{2''}
\end{equation}
since $ q^au_a = 0 $ and the heat is assumed to flow in the
radial direction. The fluid collapse rate $\Theta = u^a_{;a}$ of the
stellar model is given by
\begin{equation}
\Theta = 3\frac{\dot{B}}{AB} \label{2'''}
\end{equation}
For the line element (\ref{g1}) the Einstein field equations yield
\begin{eqnarray}\label{g3}
\mu &=& 3\frac{1}{A^2}\frac{{\dot{B}}^2}{B^2} - \frac{1}{B^2}
\left( 2\frac{B''}{B} - \frac{{B'}^2}{B^2} +
\frac{4}{r}\frac{B'}{B} \right) \,\label{g3a} \\ \nonumber \\
p &=& \frac{1}{A^2} \left(-2\frac{\ddot{B}}{B} -
\frac{{\dot{B}}^2}{B^2} +
2\frac{\dot{A}}{A}\frac{\dot{B}}{B} \right) \nonumber \\  \nonumber  \\
&& + \frac{1}{B^2} \left(\frac{{B'}^2}{B^2} +
2\frac{A'}{A}\frac{B'}{B} + \frac{2}{r}\frac{A'}{A} +
\frac{2}{r}\frac{B'}{B} \right) \, \label{g3b}  \\  \nonumber \\
p &=& -2\frac{1}{A^2}\frac{\ddot{B}}{B} +
2\frac{\dot{A}}{A^3}\frac{\dot{B}}{B} -
\frac{1}{A^2}\frac{{\dot{B}}^2}{B^2} +
\frac{1}{r}\frac{A'}{A}\frac{1}{B^2} \nonumber \\ \nonumber \\
&& +  \frac{1}{r}\frac{B'}{B^3} + \frac{A''}{A}\frac{1}{B^2} -
\frac{{B'}^2}{B^4} + \frac{B''}{B^3}
\label{g3c}  \\ \nonumber \\
q &=& -\frac{2}{AB} \left(-\frac{\dot{B'}}{B} +
\frac{B'\dot{B}}{B^2} + \frac{A'}{A}\frac{\dot{B}}{B} \right)\,.
\label{g3d}
\end{eqnarray}
where $q = (q^aq_a)^{1/2}$ is the covariant scalar measure of the heat flux. It follows that if $A(t,r)$ and $B(t,r)$
are known, then the system (\ref{g3a})--(\ref{g3d}) yield the matter variables $\mu, p$
and $q$. The nonvanishing components of the Weyl tensor are all proportional to $C_{2323}$
where
\begin{equation}
C_{2323} = \frac{r^4}{3}B^2\sin^2\theta\left[\left(\frac{A'}{A} -
\frac{B'}{B}\right)\left(\frac{1}{r} + 2\frac{B'}{B}\right) -
\left(\frac{A''}{A} - \frac{B''}{B}\right)\right].
\end{equation}
Various approaches have been utilised to generate exact solutions of the Einstein field equations (\ref{g3a}) - (\ref{g3d}). These include imposing an equation of state, acceleration-free collapse, expansion-free collapse and conformal flatness. Conformally flat, radiating spheres have been extensively studied in the literature in which the interior matter distribution is a perfect fluid with heat flux\cite{wang1,wang2,wang3}. Here we extend the conformally flat radiating solutions obtained in earlier treatments to include a null string fluid. To this end we require the vanishing of the Weyl stresses within the stellar interior which implies $C_{2323} = 0$ yielding
\emph{\emph{\emph{}}}\begin{equation}
\label{g4} A= (C_1(t)r^2+1)B.
\end{equation}
The condition of pressure isotropy is obtained by equating (\ref{g3b}) and (\ref{g3c})
\begin{equation}
\label{g5} \frac{B''}{B'}-2 \frac{B'}{B}-\frac{1}{r}=0.
\end{equation}
which easily integrates to
\begin{equation}
\label{g6} B=\frac{1}{C_2(t) r^2 +C_3(t)}.
\end{equation}
The Einstein field equations (\ref{g3a})-(\ref{g3d}) can now be written as \cite{herrr1}
\begin{eqnarray} \label{field}
\mu &=& 3\left(\frac{{\dot C_2}r^2 + {\dot C_3}}{C_1r^2 + 1}\right)^2 + 12C_2C_3 \label{fielda}\\
p &=& \frac{1}{(C_1r^2 + 1)^2}\left[2({\ddot C_2}r^2 + {\ddot C_3})(C_2r^2 + C_3)\right.\nonumber\\
&&\left.-3({\dot C_2}r^2 + {\dot C_3})^2 - 2\frac{\dot C_1}{C_1r^2 + 1}({\dot C_2}r^2 + {\dot C_3})\right.\nonumber\\
&&\left.\times (C_2r^2 + C_3)r^2\right] + \frac{4}{C_1r^2 + 1}\nonumber\\
&&\times[C_2(C_2 - 2C_1C_3)r^2 + C_3(C_1C_3 - 2C_2)]\label{fieldb}\\
q &=& 4({\dot C_3}C_1 - {\dot C_2})\left(\frac{C_2r^2 + C_3}{C_1r^2 + 1}\right)^2r\label{fieldc}\end{eqnarray}
We note from (\ref{g4}) and (\ref{g6}) that the temporal behaviour of the metric functions are yet to determined. This will be taken up in the next section. The conformally flat radiating models with the standard Vaidya atmosphere comprising pure null radiation were investigated by Herrera {\it al} \cite{herrr1} in which they established the link between inhomogeneity and dissipation. Further studies yielded new exact solutions in which the Weyl stresses vanished within the stellar interior \cite{suni,herrr2}. These models made it possible to study relaxational effects on the temperature profiles of stars with conformally flat interiors.

\section{Junction Conditions}

Here we are modeling a spherically symmetric star radiating energy in the form of a radial heat flux. This is an idealisation using the fluid approximation. In realistic stellar models gravitational collapse is preceded by an epoch of neutrino emission. The angular momentum associated with these neutrinos as they escape from the neutrino-sphere leads to tangential stresses. In addition, neutrino trapping within the stellar interior may render the core viscous. Shear viscosity has been extensively investigated in various scenarios in which the principle stresses are unequal. Our model of dissipative collapse is, in this sense, highly simplified but nevertheless serves as an important tool to test more realistic scenarios and numerical models.  The exterior spacetime is taken to the generalised Vaidya solution which represents a two-fluid atmosphere made up of null radiation and a string fluid \cite{glass1}
\begin{equation} \label{vaidya}
ds^2 = - \left(1 - 2\frac{ m(v,\sf{r})}{\sf{r}}\right)dv^2 -2
dvd{\sf{r}} + {\sf{r^2}} (d\theta^2 +\sin^2\theta
d\phi^2)
\end{equation}
where $m(v, \sf{r})$ is the mass function which represents the gravitational energy within a sphere of radius $\sf{r}$. The inclusion of strings in astrophysical scenarios has been linked to atmospheres surrounding black holes as well as dark matter constituents in globular clusters. The radial pressure within the Vaidya atmosphere can be attributed to the string tension. In the standard Vaidya envelope with photons carrying energy away from the stellar core, Santos\cite{Santos} was able to show that the radial pressure is proportional to the heat flux at the boundary of the radiating star. This boundary condition determines the temporal evolution of the model. With the generalised Vaidya atmosphere it is the photons that carry energy to the exterior spacetime with the strings diffusing inwards. Our intention is to determine the effect of the string flux on physical parameters such as temperature and luminosity in our models.
The matter source term which generates (\ref{vaidya}) via the Einstein field equations is given by
\begin{eqnarray}\label{p1extmatten}
T_{ab}^+ &=& T_{ab}^{(n)}+T_{ab}^{(m)}\label{exmatten1}\\
\nonumber\\
T_{ab}^{(n)} &=& \tilde{\mu} l_{a}l_{b}\label{exmatten2}\\
\nonumber\\
T_{ab}^{(m)} &=&
\left(\rho+P\right)\left(l_{a}n_{b}+l_{b}n_{a}\right)+Pg_{ab}\label{exmetten3}
\end{eqnarray}
 which represents a superposition of a pressureless null dust and a
null string fluid.
The two null vectors $l_a$ and $n_a$ are defined as
\begin{eqnarray}\label{exnullvecs}
l_{a} &=& \delta_{a}^0 \label{exnullvec1}\\
\nonumber\\
n_{a} &=& \frac{1}{2}\left[1-2\frac{m(v,
\sf{r})}{\sf{r}}\right]\delta_{a}^{0}+\delta_{a}^{1}\label{exnullvec2}
\end{eqnarray}
where $l_{a}l^{a}=n_{a}n^{a}=0$ and $l_{a}n^{a}=-1$. The matter variables for the exterior are generated via the Einstein
field equations $G_{ab}^+=T_{ab}^+$ which yield
\begin{eqnarray}\label{exeinfield}
\epsilon &=& -2\frac{m_v}{\sf{r}^2}\tilde{v}^2\label{exeinfield1}\\
\nonumber\\
\rho &=& 2\frac{m_{\sf{r}}}{\sf{r}^2}\label{exeinfield2a}\\
\nonumber\\
P &=& -\frac{m_{\sf{r}\sf{r}}}{\sf{r}}\label{exeinfield3}
\end{eqnarray}
where subscripts $v$ and ${\sf r}$ abbreviate to $\partial/\partial{v}$ and $\partial/\partial{\sf r}$ respectively. The energy density of the null radiation is denoted by $\epsilon$, $\rho$ is the null string density and $P$ is the null string pressure. In the case of $\rho = P = 0$, we regain the standard Vaidya solution with $m = m(v)$.

In order to generate a complete model of a radiating star the line element (\ref{g1}) must be smoothly matched to the exterior spacetime (\ref{vaidya}) across a time-like hypersurface. The matching conditions were first obtained by Maharaj {\em et al} \cite{Sunil} and we state the main results here.
The mass profile of the collapsing sphere is given by
\begin{equation}
m(v, {\sf{r}})= \left[\frac{rB}{2}\left(1+r^2 \frac{\dot{B}^2}{A^2}
-
\frac{1}{B^2}(B+rB^{\prime})^2\right)\right]_{\Sigma}\label{stdmass}
\end{equation}
 which is the total gravitational energy contained within the
stellar surface $\Sigma$. The continuity of the momentum flux across the boundary of the star yields
\begin{equation}
p=\left[qB-\rho\right]_{\Sigma}\label{densres}
\end{equation}
which generalises the results obtained by Santos in 1985. It is clear from (\ref{densres}) that
the pressure at the boundary of the collapsing star depends on the magnitude of the heat flux $q$ and the exterior
string density $\rho$. For our line element
(\ref{g1}) and the assumption of vanishing Weyl stresses,
(\ref{densres}) reduces to the nonlinear equation
\begin{eqnarray}
&&
\ddot{C_2}b^2+\ddot{C_3}-\frac{3}{2}\frac{(\dot{C_2}b^2+\dot{C_3})^2}{C_2b^2+C_3}
-\frac{\dot{C_1}b^2(\dot{C_2}b^2+\dot{C_3})}{C_1b^2+1}-2(C_1
\dot{C_3}-\dot{C_2})b \nonumber\\
&&
+2\frac{(C_1b^2+1)}{C_2b^2+C_3}[C_2(C_2-2C_1C_3)b^2+C_3(C_1C_3-2C_2)]+
\frac{\rho}{2} \frac{(C_1b^2+1)^2}{C_2b^2+C_3}=0, \nonumber \\
&& \label{g9}
\end{eqnarray}
where $r=b$ determines the boundary of the star. If we assume a constant string density ($\rho = constant$), then we are in a position to obtain
the following exact solution
\begin{eqnarray}\label{g15}
C_1 &=& \frac{4 \alpha}{4C_3(2\alpha-3C_3)-\rho
b^2}\left[\frac{\alpha}{b^2}-\frac{4C_3}{b^2}+\frac{3C_3^2}{\alpha
b^2} -\frac{\dot{C_3}}{b}+\frac{\rho}{4 \alpha}\right], U\neq
0
 \label{g15a}\\
C_2&=& \frac{\alpha -C_3}{b^2} \label{g15b}\\
C_3 &=& \mbox{arbitrary function of}~ t \label{g15c}
\end{eqnarray}
where $\alpha$ is an arbitrary constant and we have defined $U(t) = C_1(t)b^2+1$. It is interesting to note that the case of constant string density is mathematically equivalent to the radiating models studied by Thirukannesh {\em et al} \cite{Moops} in which they considered dissipative collapse in the presence of the cosmological constant.
Furthermore, we note that in the case of vanishing string density, ie., $m = m(v)$, the boundary condition (\ref{g15}) reduces to the problem investigated by Rajah {\em et al} \cite{suri1} in which the exterior spacetime was taken to be the standard Vaidya solution. We will now utilise (\ref{g15a})-(\ref{g15c}) to investigate the effect of dissipation on the collapse process, more importantly, the influence of the string density on the thermal behaviour of our model.

\section{Relaxational effects}

It has been shown that relaxational effects give rise to substantially different luminosity profiles as compared to profiles corresponding to vanishing relaxational times. If we consider the early epoch of stellar collapse, the core temperature could be as high as $T \approx 10^{9}$K in which case the relaxation time is of the order of $\tau \approx 10^{-4}$s. For a temperature of $T \approx 10^6$K, the relaxation time is of the order of $\tau \approx 10^2$s \cite{dp1,dp2}.
In order to study the role played by the relaxation time when the star leaves hydrostatic equilibrium we utilise a causal heat transport equation of Maxwell-Cattaneo form given by \cite{and1}
\begin{equation}{\tau} h_a{}^b {\dot q}_{b} + q_a =
-\kappa(h_a{}^b \nabla_b T + T {\dot u}_a) \label{cmc}
\end{equation}
where we have assumed no viscous-heat coupling of the thermodynamical fluxes. Various authors have highlighted the importance of the relaxation time on the evolution of the temperature and luminosity profiles of the collapsing star. The Eckart tranport equation obtained by setting $\tau = 0$ in (\ref{cmc}) suffers various pathologies such as the prediction of infinite propagation velocities for the thermal signals and unstable equilibrium states \cite{ani}. A general framework adopting a full causal approach to dissipative collapse was presented by Herrera {\it et al} in which the full causal transport equations are coupled to the dynamical equations governing the stellar interior \cite{herrr3}. We assume that during late stages of collapse heat dissipation occurs via neutrino emission. The neutrinos are thermally generated within the stellar core with energies of the order of $k_{B}T$. At neutron star densities neutrino trapping takes place via electron-neutrino scattering and nucleon absorption. The mean collision time for thermally generated neutrinos is given by
\begin{equation} \label{tauu} \tau_{\rm c} \propto T^{-3/2}
\end{equation} to good approximation \cite{marti}. Following (\ref{tauu}) we adopt a power-law dependence for the thermal conductivity and relaxation time:
\begin{equation}
\kappa =\gamma T^3{\tau}_{\rm c}, \hspace{2cm} \tau_{\rm c}
=\left({\alpha\over\gamma}\right) T^{-\sigma} \label{a28}\,
\end{equation}
where $\alpha \geq 0$, $\gamma \geq 0$ and $\sigma \geq 0$ are
constants.
We further assume that the relaxation time is directly proportional to the mean collision time \begin{equation}
\tau =\left({\beta \gamma \over \alpha}\right) \tau_{\rm c}
\label{a30}\,
\end{equation}
where $\beta$ ($\geq 0$) is a constant. The
causal transport equation (\ref{cmc}) together with the above assumptions reduces to
\begin{equation}
\beta (qB)^{\dot{}} T^{-\sigma} + A (q B) = - \alpha
\frac{T^{3-\sigma} (AT)'}{B} \label{temp1} \,.
\end{equation} where the Eckart temperature $T_0$ is obtained by setting $\beta = 0$ in (\ref{temp1}). We are in a position to study the evolution of the temperature profile for the case $\sigma = 0$ which corresponds to constant collision time.
In the case of constant mean collision time the causal transport equation (\ref{temp1}) is simply integrated to
yield,
\begin{equation} (AT)^4 = - \frac{4}{\alpha} \left[\beta\int A^3 B
(qB)_{,t}{\mathrm d} r + \int A^4 q B^2 {\mathrm d} r\right] +
F(t) \label{caus0} \end{equation} where $F(t)$ is an arbitrary function of integration. The function ${F}(t)$ can be determined from the effective surface temperature of a star as given by\cite{Bonnor}
\begin{equation} \label{f10}
\left({T^4}\right)_{\Sigma} =
\left(\frac{1}{r^2B^2}\right)_{\Sigma}\left({L_{\infty}\over 4\pi\delta}
\right)
\end{equation}
where $L_{\infty}$ is the total luminosity at
infinity and $\delta$ ($>0$) is a constant.
We make use of solution
(\ref{g15a})-(\ref{g15c}) for the case $U \neq 0$ and $C_3(t) = a
e^{t}$ where $a$ is a constant and $-\infty \leq t \leq 0$. The choice of $C_3(t)$ ensures that the matter variables as defined by (\ref{fielda})-(\ref{fieldc}) and the metric functions are physically viable for some finite epoch of the collapse period. In order to generate the plots in figures 1 and 2 we choose the following values for our arbitrary constants: $a = 1, \alpha = 1$, $\rho = 10^2$ and $b = 10$. The causal temperature profiles correspond to $\beta = 0$ and an earlier time $t = -100$ (stellar fluid is close to hydrostatic equilibrium). For the causal temperature we take $\beta = 10^8$ corresponding to late-time evolution, $t = -1$ (fluid is in quasi-static equilibrium). In figure 1 we have plotted the noncausal temperature profiles ($\beta = 0$), for ($\rho = 0$)  corresponding to the standard Vaidya atmosphere composed purely of null radiation and $\rho = {\mbox constant}$  representing a two-fluid atmosphere. In this limit the two temperature profiles are very similar in magnitude and behaviour throughout the stellar core. This is expected as the stellar fluid is very close to hydrostatic equilibrium, corresponding to early stages of collapse. The noncausal temperature at the center of the star is enhanced by approximately 15$\%$ in the presence of the string fluid in the exterior. Figure 2 clearly indicates that the casual temperatures are higher than their noncausal counterparts throughout the star. The causal temperature is sensitive to the presence of the string distribution in the exterior and our model indicates a $20 \%$ to $30\%$ enhancement of the temperature at $r = 0$. It is also evident that the temperature for the two-fluid atmosphere diverges significantly from the temperature of the standard Vaidya exterior. We can think of this late-time evolution of the temperature in terms of the heat generated during this epoch. As the core collapses the heat generated will be higher than the initial stages of collapse. To maintain a higher temperature distribution throughout the interior of the star, we can think of the string distribution in the exterior as reducing the heat dissipation from the interior.

\section{Horizon formation}

The total luminosity for an observer at rest at infinity is given by \cite{pin}
\begin{equation}  \label{b25}
L_{\infty}(v)  =  -\left(\displaystyle\frac{dm}{dv}\right)_{\Sigma}
\end{equation}
where $\frac{dm}{dv} \leq 0$ since $L_{\infty}$ is positive.
An observer with four--velocity $v^a = (\dot{v}, \dot{\sf{r}}, 0, 0)$ located
on $\Sigma$ has proper time $ \tau $ related to the time
$ t $ by $ d \tau = Adt $.
The radiation energy density that this observer measures on $
\Sigma $ is
\[
{\epsilon}_{\Sigma} =
\frac{1}{4\pi}\left(-\displaystyle\frac{{\dot{v}}^2}{{\sf
{r}}^2}
\displaystyle\frac{dm}{dv} \right)_{\Sigma}
\]
and the luminosity observed on $ \Sigma$ can be written as
\[
L_{\Sigma} = 4{\pi}{\sf{r}}^2{\epsilon}_{\Sigma}
\]
The boundary redshift $z_{\Sigma}$ of the radiation
emitted by the star is given by
\begin{equation}  \label{b26}
1 + z_{\Sigma} = \left(\displaystyle\frac{dv}{d \tau}\right)_{\Sigma} = \left(\frac{R'}{B} + \frac{{\dot R}}{A}\right)_{\Sigma}^{-1}
\end{equation}
which can be used to determine the time of formation of the horizon.
The above expressions allow us to write
\[
1 + z_{\Sigma} = \left(\frac{L_{\Sigma}}{L_{\infty}}\right)^{\frac{1}{2}}
\]
which relates the luminosities $L_{\Sigma}$ to $L_{\infty}$
via the surface redshift.
For our model under consideration the surface luminosity is given by
\begin{equation} \label{sur}
L_{\infty} = \frac{ae^t(-2ae^t - \alpha)^2(4ae^t(3ae^t - 2\alpha)+ b^2\rho)}{2b\alpha^2(a(2 + b)e^t - \alpha)}
\end{equation}
In order to determine the time of horizon formation, the time that no radiation is observed by an observer at infinity we require
\[
L_{\infty} = 0\]
This requirement yields three roots
\begin{eqnarray}
  t_1 &=& \ln{\left[\frac{2a\alpha - \sqrt{4a^2\alpha^2 - 3a^2b^2\rho}}{6a^2}\right]}\\
   t_2 &=& \ln{\left[\frac{2a\alpha + \sqrt{4a^2\alpha^2 - 3a^2b^2\rho}}{6a^2}\right]}\\
   t_3 &=& \ln{\left[\frac{\alpha}{2a}\right]}
\end{eqnarray}
Let us consider an atmosphere consisting solely of null radiation ($\rho = 0$). This gives $t_1 = -\infty$, the time of onset of collapse and $t_2 = \ln{\left[\frac{2\alpha}{3a}\right]}$ which is the time of formation of the horizon. We note that $t_3 = \ln{\left[\frac{\alpha}{2a}\right]} > t_2$ occurs after the horizon forms, ie. the collapse has already proceeded to a black hole state. The picture is altered when $\rho$ is nonzero. The onset of collapse occurs at ${\tilde t}_1$ (observations of the collapse begins) and proceeds to form an horizon at ${\tilde t}_2 \leq t_2$. This means that the horizon forms earlier in the presence of the two-fluid atmosphere. Recall that our analyses of both the causal and noncausal temperature profiles indicate a higher core temperature in the presence of a two-fluid atmosphere ($\rho \neq 0$). This indicates that the heat dissipation to the exterior is reduced which has the corresponding effect of reducing the outward pressure required to maintain stability. This diminishing of the outward pressure in the presence of the string density results in a greater collapse rate, thus resulting in the horizon forming earlier. From (\ref{sur}) we note that $L_\infty$ diverges at \[
t_{div} = \ln\left[\frac{\alpha}{a(2 + b)}\right]\] The singular behaviour of the luminosity implies that the energy per unit area of the stellar surface per unit time diverges thus implying that our model breaks down here. In the case of a pure Vaidya atmosphere we note that $t_{div} < t_2$ where $t_2$ is the time of formation of the horizon. This means that the luminosity diverges before the time of formation of the horizon which restricts the evolution of our model. In the case of the generalised Vaidya atmosphere, the string component can lead to an earlier formation of the horizon thus avoiding the divergence of the luminosity.

\section{Conclusion}
We have modeled a radiating star undergoing dissipative collapse with the exterior spacetime described by the generalised Vaidya solution. One can conceive of the scenario as a spherically symmetric matter distribution collapsing and dissipating energy in the form of a radial heat flux while surrounded by a two-fluid atmosphere composed of null radiation and a null string fluid. We have solved the boundary condition required for the smooth matching of the interior spacetime to the generalised Vaidya solution by assuming a constant string density for the surrounding atmosphere. The thermal behaviour of a particular model is studied by employing a heat transport equation of Maxwell-Cattaneo form. We have shown that during the initial stages of collapse the influence of the string density is minimal on the temperature profile. During late-time collapse the influence of the string density is much more pronounced, leading to very different temperature profiles within the stellar core. It would be interesting to investigate the influence of the string density on the temperature evolution for a more general matter profile, ie., $\rho = \rho(v,\sf{r})$.

\section*{Acknowledgments}
The author is grateful to Professor L. Herrera, Universidad del Pai$^{\prime}$s Vasco for useful comments and enlightening suggestions which improved the presentation of the results contained in this manuscript. The author acknowledges valuable suggestions and insightful recommendations made by the anonymous referee which helped clarify the main results in this paper.

\newpage
\begin{figure}
\centering
\includegraphics[scale=0.6]{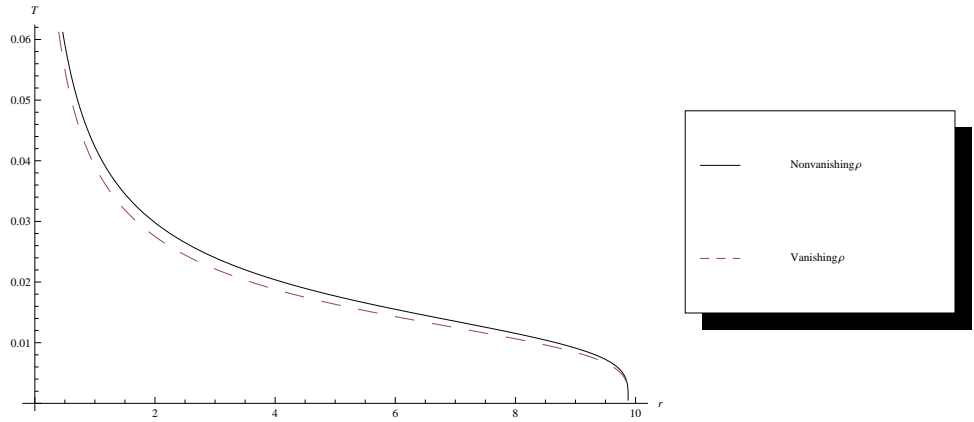}\caption{Noncausal temperature profiles} \label{fig1}
\end{figure}

\begin{figure}
\centering
\includegraphics[scale=0.6]{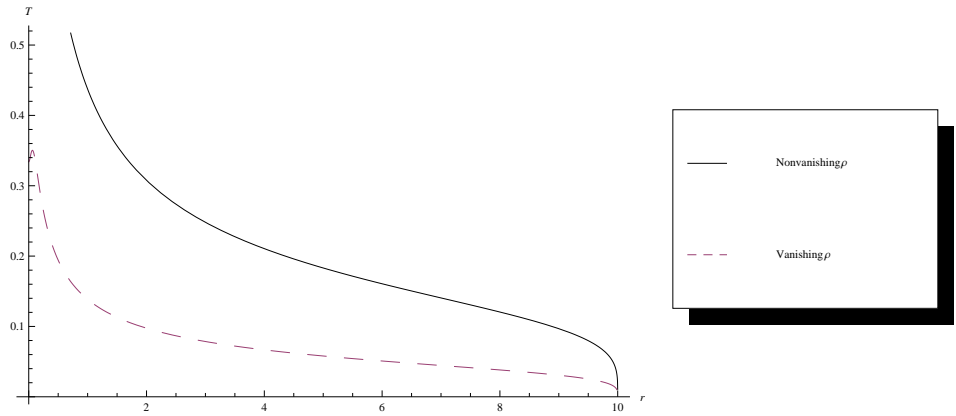}\caption{Causal temperature profiles} \label{fig2}
\end{figure}
\end{document}